\documentclass[journal]{IEEEtran}
\usepackage{graphicx}
\usepackage{cite}
\usepackage{picinpar}
\usepackage{amsmath}
\usepackage{stfloats}
\usepackage{url}
\usepackage{flushend}
\usepackage{colortbl}
\usepackage{soul}
\usepackage{multirow}
\usepackage{pifont}
\usepackage{color}
\usepackage{alltt}
\usepackage{enumerate}
\usepackage{siunitx}
\usepackage{epstopdf}
\usepackage{pbox}
\usepackage[official]{eurosym}
\usepackage[T1]{fontenc}
\usepackage[utf8]{inputenc}
\usepackage{graphicx}
\usepackage{etoolbox}
\usepackage[breaklinks=true,colorlinks=true,linkcolor=blue,urlcolor=blue,citecolor=blue]{hyperref}
\usepackage{upgreek}
\usepackage{braket}
\usepackage[utf8]{inputenc}
\usepackage[english]{babel}
\usepackage{cite}
\begin{document}

\title{1.6 {GHz} Frequency Scanning of a 482~nm Laser Stabilized using Electromagnetically Induced Transparency}

\author{Krishnapriya Subramonian Rajasree, Kristoffer Karlsson, Tridib Ray,
        and S\'ile {Nic Chormaic} 
\thanks{Manuscript received xxx This work was supported by OIST Graduate University and JSPS Grant-in-Aid for Scientific Research (C) Grant Number 19K05316.}
\thanks{K.P. Subramonian Rajasree, K. Karlsson and S. Nic Chormaic are with the Light-Matter Interactions for Quantum Technologies Unit, Okinawa Institute of Science and Technology Graduate University, Onna, Okinawa 904-0495, Japan (e-mail: sile.nicchormaic@oist.jp).   }
\thanks{T. Ray is with Laboratoire Kastler Brossel, Sorbonne Universit\'e, CNRS, ENS-Universit\'e PSL, Coll\`ege de France, 4 place Jussieu, 75005 Paris, France. }}


\title{1.6 {GHz} Frequency Scanning of a 482~nm Laser Stabilized using Electromagnetically Induced Transparency}

\maketitle

\begin{abstract}
We propose a  method to continuously frequency shift a target laser that is frequency stabilized by a reference laser, which is several hundreds of nanometers detuned.  We demonstrate the technique using the $\mathbf{5S_{1/2}} \rightarrow \mathbf{5P_{3/2}} \rightarrow \mathbf{29D_{5/2}}$ Rydberg transition in $^{87}$Rb vapor and lock the 482~nm target laser to the 780~nm reference laser using the cascaded electromagnetically induced transparency signal. The stabilized frequency of the target laser can be shifted by about 1.6 GHz by phase modulating the reference laser using a waveguide-type electro-optical modulator. This simple method for stable frequency shifting can be used in atomic or molecular physics experiments that require a laser frequency scanning range on the order of several GHz.
\end{abstract}


\IEEEpeerreviewmaketitle

\section{Introduction}

\IEEEPARstart{M}{any} modern atomic and molecular physics experiments require a laser with a stabilized frequency, which can be shifted or scanned over a desired amount, typically up to a few GHz, in order to address a specific transition between energy levels of interest. Usually, resonant absorption on optical transitions, frequency combs, or optical cavity resonances are used as a reference to stabilize lasers to a particular frequency.  Some techniques in common use for direct frequency referencing  include saturated absorption spectroscopy \cite{Preston_1996}, Sagnac interferometry \cite{Jundt_2003}, Pound-Drever-Hall locking \cite{Drever_83,Hirata_2014}, and dichroic atomic vapor laser locking \cite{Becerra_2009}. In addition, techniques such as electromagnetically induced transparency  (EIT) or beat note locking can be used to lock the relative frequency of two lasers \cite{Boller_1991, Lukin_98, Flei_2005, Bell_2007, Islam_14}. The first method is widely employed in situations where it is difficult to obtain a direct absorption signal, such as the excitation of a neutral atom to a Rydberg level using a cascaded process \cite{Abel_2009,Langbecker_2017, Jia_2020}. The disadvantage of this technique is that it tends to provide discrete frequency reference shifts. Beat note locking can be used for a continuously varying frequency shift, but the target laser can only be separated from the reference laser by up to a few GHz.   

Some of the more common methods to arbitrarily shift the laser frequency from a stabilized reference point  include using an acousto-optical modulator (AOM) or  an electro-optical modulator (EOM). Laser stabilization using an AOM by modulating the carrier  frequency has  been demonstrated \cite{Aldous_17} and this method is suitable for probing very narrow absorption features. To reach GHz shifts using an AOM, a single pass is typically not sufficient. Therefore, configurations including 2-, 3-, 4-, 6-, and even 12-passes \cite{Donley_2005,de_2012,lu_2017,buch_2000,Chao_2020} have been implemented.     When using an AOM, the frequency-shifted light is separated from the carrier (that is the zeroth order beam) via diffraction. However, the entire frequency range under consideration cannot be covered by a single modulator. Moreover, the change in frequency can cause a change in intensity of the frequency-shifted light if the diffraction efficiency of the modulator is not uniform over the full frequency range. 

In contrast, when using an EOM, the attainable frequency shift can be much larger (for the non-resonant or broadband case); up to 40~GHz offset using a 10~GHz modulator has been demonstrated \cite{Peng_2014} by using the 4th-order sidebands and, more recently, up to 46~GHz was  obtained using the 10th-order sidebands \cite{Harada_16}. However, EOMs require a high driving voltage over the  desired range and the shifted sidebands co-propagate with the carrier, meaning they cannot be spatially separated unless a narrow-band cavity is used {\cite{Gatti:15}.

In this letter,  we demonstrate  laser locking and subsequent frequency shifting up to $\pm 800$~MHz of a 482~nm target laser using EIT with a waveguide-type EOM employed for shifting the frequency of a 780~nm reference laser. The quality of the frequency stabilization achieved is demonstrated in terms of both stability and scanning range. The work was motivated by our need for a tunable, frequency-stabilized laser for cascaded Rydberg atom excitation in $^{87}$Rb \cite{rajasree_prrea_2020},

\section{Experimental Details}
We focused on frequency shifting a frequency stabilized laser so as to excite $^{87}$Rb atoms from the  $5S_{1/2}$ ground level to the $29D_{5/2}$ Rydberg level via the $5P_{3/2}$ intermediate level using a cascaded excitation process, as shown in Fig. \ref{REITdia}(a).  The experimental setup is shown in Fig.~\ref{REITdia}(b).  We used a natural abundance rubidium vapor cell to frequency lock a 780~nm laser (DL pro, Toptica) on the $^{85}$Rb $5S_{1/2}(\mathrm{F}= 3) \rightarrow 5P_{3/2}(\mathrm{F}^\prime = 3,4)_{co}$ transition using saturated absorption spectroscopy (SAS). The locked 780~nm laser acted as the reference while the target laser at 482~nm was derived from a frequency-doubled high power laser (TA SHG pro, Toptica). The aim was to scan the target laser across the Rydberg transition (see Fig.~\ref{REITdia}(a)), for which the direct absorption strength is very weak, hence a signal is difficult to detect. 
 
 \begin{figure}[ht]
        \centering
        \includegraphics[width=3 in]{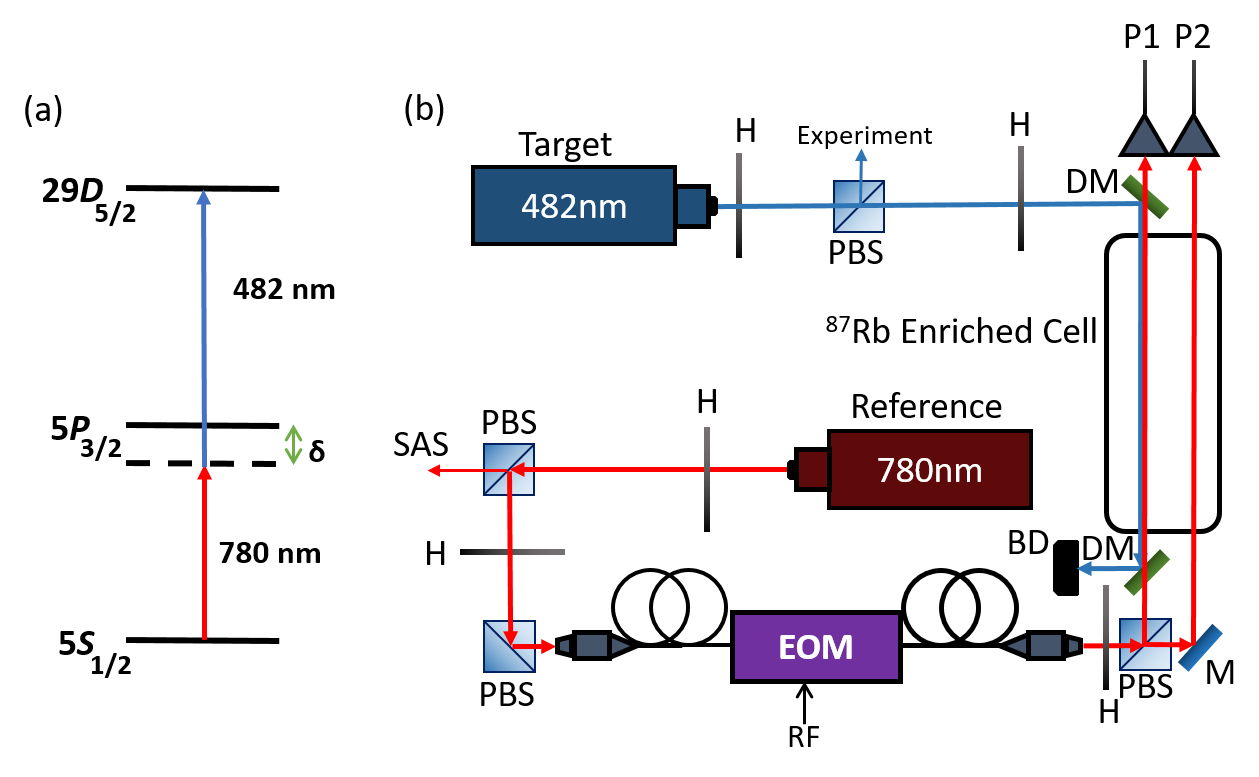}
        \caption{{(a) Simplified energy level diagram for $^{87}$Rb showing the relevant transitions.  The ground level $5{S}_{1/2}$, the intermediate level $5P_{3/2}$ and the Rydberg level  $29D_{5/2}$ constitute the cascaded three-level system.  $\delta$ is the frequency shift on the 780~nm reference laser from the resonance condition.  (b) Schematic of the experimental setup. The 482~nm target laser and the 780~nm reference laser pass through the $^{87}$Rb-enriched vapor cell in a counter-propagating configuration. Absorption of the 780~nm light in the absence and presence of the 482~nm light is detected on P1 and P2, respectively. The difference between the two signals yields the Doppler-free EIT signal, which is used  to frequency stabilize the 482~nm laser. EOM: electro-optic modulator; H: half-wave plate; PBS: polarizing beam splitter; M: mirror; DM: dichroic mirror; BD: Beam dump; P1, P2: balanced photodiodes. The different colored arrows indicate different wavelengths of light, red for 780~nm and blue for 482~nm.}}
        \label{REITdia}
    \end{figure}

 For stabilizing the target laser frequency we relied on Rydberg EIT \cite{mohapatra_2007}; the 780~nm reference laser at a low power (50 $\mu$W) was used as the probe and the 482~nm target laser (90~mW) acted as the pump to produce the EIT signal. The Rydberg EIT experiments were done in a $^{87}$Rb-enriched vapor cell (TT-RB87-75-V-P, TRIAD Technology Inc.) of dimensions 25~mm $\times$ 75~mm, at room temperature. Since  the reference laser addressed the $^{85}$Rb $5S_{1/2}(F= 3) \rightarrow 5P_{3/2}(F^\prime = 3,4)_{co}$ transition and the vapor cell was enriched with the $^{87}$Rb isotope, the conditions necessary to observe EIT were not met. To overcome this, we sent the reference laser  through a waveguide-type EOM (NIR-NPX800 LN-10, Photline Technologies) to produce sidebands at the desired frequency. The frequency separation between the  $^{85}$Rb $5S_{1/2}(F= 3)$ $\rightarrow$ $5P_{3/2}(F^\prime = 3,4)_{co}$ and the $^{87}$Rb $5S_{1/2}(F=2)$ $\rightarrow$ $5P_{3/2}(F^\prime = 3)$ transition, used to drive the Rydberg excitation, is 1.0662~GHz and $\delta$ represents a shift from this frequency (see Figs. \ref{REITdia}(a) and \ref{level}(a)). 
 
The radio frequency (RF) signal to the EOM was chosen so as to adjust the sideband frequency by (1.0662$\pm \delta$)~GHz, thereby ensuring that the EOM output satisfied the EIT condition in $^{87}$Rb.  This then guaranteed that the frequency of the 482~nm laser was also correct. The transmission of the 780~nm probe laser through the atomic vapor in the presence or absence of the 482~nm pump was detected using photodiodes, P1 and P2, respectively, see Fig.\ref{REITdia}(b). When the 482~nm laser was resonant with the Rydberg transition, we observed a peak in the 780~nm transmission and this was the desired EIT signal, see Fig. \ref{level}(b). Here, in this EIT-based locking technique, the EIT signal was generated from a two-photon resonance and, as such, the target laser was locked relative to the reference laser. Only one of the EOM sidebands participated in the EIT process and both the carrier and the other sideband passed through the vapor cell without any interaction. The  EIT signal was modulated so as to yield an error signal, which was then used to lock the target laser.
 
  \begin{figure}[ht]
        \centering
        \includegraphics[width=3.5 in]{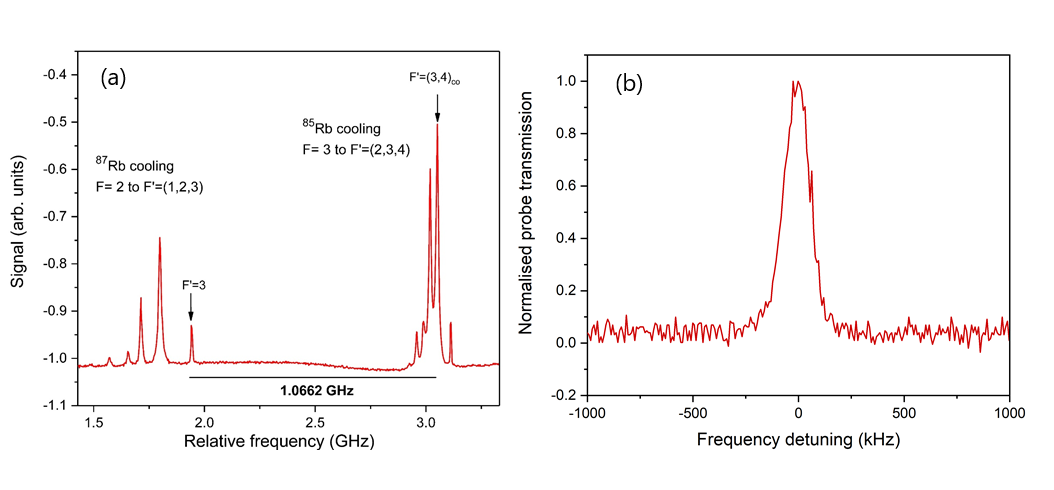}
        \caption{(a) Saturated absorption spectrum for Rb obtained from a commercial frequency locking interface (Digilock, Toptica). The frequency separation between the  $^{85}$Rb $5S_{1/2}(F= 3)$ $\rightarrow$ $5P_{3/2}(F^\prime = 3,4)_{co}$ and the $^{87}$Rb $5S_{1/2}(F=2)$ $\rightarrow$ $5P_{3/2}(F^\prime = 3)$ transition is 1.0662~GHz as shown. (b) A typical 780~nm probe EIT signal used for locking the 482~nm target laser to the $^{87}$Rb $5P_{3/2} \rightarrow  29D_{5/2}$ transition. }
        \label{level}
    \end{figure} 
    
 \section{Performance and Discussion}   

 The RF signal applied to the EOM was varied to ensure that the sidebands were detuned from the $^{87}$Rb cooling transition, $5S_{1/2}(F=2) \rightarrow 5P_{3/2}(F^\prime = 3)$, by $\delta$. To satisfy the resonance condition, the frequency of the target laser  shifts by an equivalent amount $-\delta$. In fact, the RF signal to the EOM could either be kept fixed so as to lock the target laser to a specific frequency or it could be varied continuously in order to shift the frequency of the target laser. Importantly, this technique does not change the intensity of the target laser output. Figure \ref{scanning} is a plot of the 482~nm target laser frequency shift as a function of the applied RF signal to the EOM.  We see that the target laser shifted over about 1600 MHz ($\pm 800$~MHz) as the RF signal was changed by 892~MHz ($\pm446$~MHz). No compensation techniques were used. \\   
     \begin{figure}[ht]
        \centering
         \includegraphics[width=3 in]{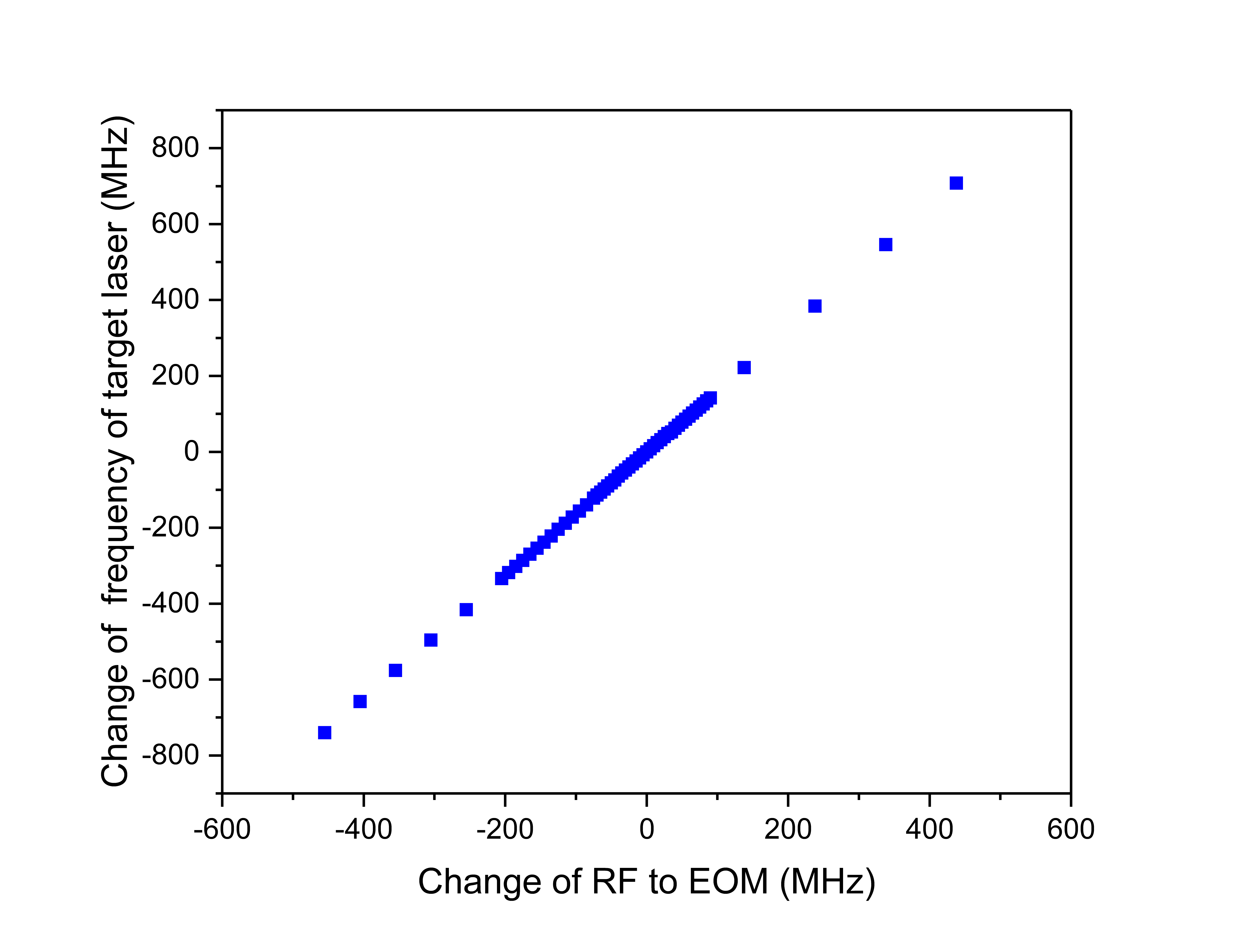}
        \caption{Shift in the frequency of the 482~nm target laser as a function of the applied radio frequency to the EOM. The total shift achievable is on the order of 1600~MHz. The zero frequency corresponds to $\delta = 0$. }
        \label{scanning}
    \end{figure}

Figure \ref{stability} presents the frequency stability of the 482~nm target laser when locked using the method presented. We measured the frequency of the 482~nm laser with a wavelength meter (WS-6, High Finesse). The frequency fluctuation was $\pm$0.2 MHz over a period of 80 minutes. The lock stability using the first order output from the EOM was comparable and equivalently as good as the zeroth order lock previously demonstrated for a Rydberg level \cite{Jiao_2016} and the frequency stability demonstrated was sufficient for a typical atomic physics experiment \cite{rajasree_prrea_2020}.    

 \begin{figure}[ht]
        \centering
        \includegraphics[width=3 in]{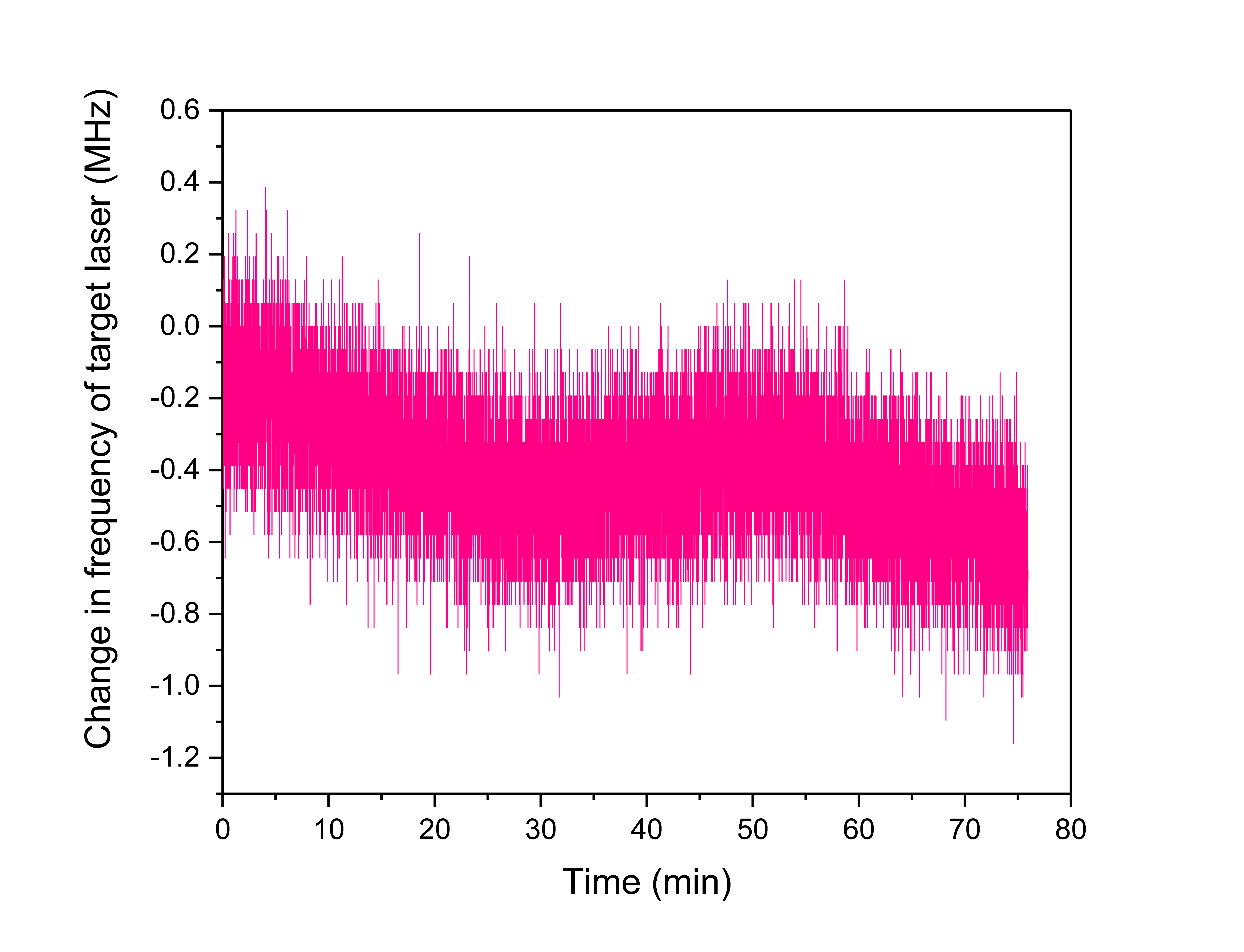}
        \caption{Fluctuations in the frequency of the 482~nm target laser as a function of time when frequency locking is on.}
        \label{stability}
    \end{figure}
The limitation on the frequency scanning range was $\pm 800$~MHz. This arises from the Doppler width of the $^{87}$Rb cooling transition manifold. Beyond the Doppler broadened absorption, it becomes harder to obtain a EIT peak. An alternative approach would be to lock the frequency of the 482~nm laser to a reference laser using an optical phase-locked loop \cite{Ferrero_08}. The beat note generated could be referenced to an RF signal and the frequency of the target laser could then be varied more or less arbitrarily without changing the light intensity \cite{Numata_2012}. However, the implementation of this technique would be significantly more complicated and limited to  reference and target lasers with adjacent frequencies, typically not separated by more than a few GHz. In contrast, our technique provides a simple, stable method to produce a phase-coherent pair of laser beams which are hundreds of nm apart.

\section{Conclusion}
In summary, we have shown a novel method to shift a stabilized laser by a desired frequency. A reference laser at 780~nm, which was hundreds of nanometres away from the target laser at 482~nm, was used for the frequency stabilization. This method of achieving a subnatural linewidth stable target laser with a long range frequency scan (1.6 GHz) could be used for atomic physics experiments involving Rydberg levels \cite{rajasree_prrea_2020}. Other than for Rydberg  experiments, this frequency locking and shifting method could provide a simple alternative when  lasers with a large detuning are required, for example,  in long-term precision measurements, such as frequency chirping, atom clocks, atom interferometers, and laser frequency modulation.

\bibliographystyle{IEEEtran}
\bibliography{articleArxiv}

\end{document}